\documentstyle[multicol,aps,epsf]{revtex}

\begin{document}
\title{Alternating Kinetics of Annihilating Random Walks Near 
a Free Interface}
\author{L.~Frachebourg$^\ast$, P.~L.~Krapivsky$\dag$, and S.~Redner$\dag$}

\address{$^\ast$Laboratoire de Physique Statistique, Ecole Normale 
Sup\'erieure, F-75231 Paris Cedex 05, France}
\address{$\dag$Center for Polymer Studies and Department of Physics,
Boston University, Boston, MA 02215}
\maketitle
\begin{abstract} 

The kinetics of annihilating random walks in one dimension, with the
half-line $x>0$ initially filled, is investigated.  The survival
probability of the $n^{\rm th}$ particle from the interface exhibits
power-law decay, $S_n(t)\sim t^{-\alpha_n}$, with $\alpha_n \approx
0.225$ for $n=1$ and {\em all} odd values of $n$; for all $n$ even, a
faster decay with $\alpha_n\approx 0.865$ is observed.  From
consideration of the eventual survival probability in a finite cluster
of particles, the rigorous bound $\alpha_1\leq 1/4$ is derived, while a
heuristic argument gives $\alpha_1\approx 3\sqrt{3}/8 = 0.2067\ldots$.
Numerically, this latter value appears to be a lower bound for
$\alpha_1$.  The average position of the first particle moves to the
right approximately as $1.7\, t^{1/2}$, with a relatively sharp and
asymmetric probability distribution.

\end{abstract}
\begin{multicols}{2}

Annihilating random walks (ARWs) represent a simple but ubiquitous
reaction process in which particles diffuse and annihilate
whenever they meet\cite{privman}.  In addition to providing
general insights about non-equilibrium phenomena, ARWs underlie a
variety of basic kinetic processes ranging from the voter
model\cite{lig} and the kinetic Ising-Glauber model\cite{p1}, to
reaction-diffusion systems\cite{p2} and wetting phenomena\cite{fisher}.
In one dimension, powerful exact solution methods have been developed to
understand many kinetic and spatial properties of
ARWs\cite{glauber,bed,fel,de96,spouge,pk}.

While much is known about ARWs under homogeneous conditions, the role of
spatial heterogeneity in such non-equilibrium systems is relatively
unexplored.  Our particular interest is to understand the influence of a
free interface on the asymptotic behavior of ARWs.  For equilibrium
systems at criticality, the presence of such a free interface gives rise
to well understood surface critical behavior which is characterized by
associated surface critical exponents\cite{diehl,leclair}.  For reactive
systems, in contrast, while there has been some progress in
understanding the role of heterogeneity in the intrinsic properties of
the reactants\cite{doering,d1,d2,d3,d4}, the role of heterogeneity in
the form of a free interface is still unexplored.

In this paper, we investigate basic properties of a one-dimensional
semi-infinite system of ARWs.  We consider a linear chain in which one
particle initially occupies each lattice site $n$ for $n>0$, while the
system is empty for $n\leq 0$.  Far from the interface, the behavior
should coincide with that of the homogeneous system.  However, the
interfacial region is a less reactive environment because one side of
the system is initially empty.  One might thus anticipate that particles
near the interface should exhibit slower kinetics and different spatial
properties than bulk particles.  This expectation is only partly
correct; in fact, every other particle near the interface exhibits
faster kinetics compared to bulk particles.  Our work further indicates
that there are only two apparently independent ``surface'' exponents
which characterize the asymptotic particle survival probabilities.  This
surface behavior eventually governs the entire system, although it
penetrates slowly into the bulk by diffusion.

To quantify the phenomena that are governed by the existence of the
interface in the semi-infinite ARW system, our work is organized around
the following basic questions:

\begin{itemize}

\item What is the probability that the first particle survives until
time $t$, $S_1(t)$?  More generally, what is the survival probability
for the $n^{\rm th}$ particle from the interface, $S_n(t)$?

\item What is the probability that particles $i$ and $j$ react as a
function of $|i-j|$ (with $i=1$ or 2 and $j$ arbitrary)?

\item What is the spatial density distribution near the interface?

\end{itemize}

To answer the first question, let us introduce the exponents $\alpha_n$
to characterize the probability that the $n^{\rm th}$ particle survives
until time $t$, $S_n(t)\sim t^{-\alpha_n}$.  We first argue that there
are only two independent exponents -- one for $n$ odd and a second for
$n$ even.  To support this assertion, it is instructive to examine
finite particle systems.  For 3 particles, the first and third particles
have a finite probability to survive indefinitely, while the second
certainly dies, with a survival probability which decays as
$t^{-3/2}$\cite{spouge,fg,dba}.  Similar asymptotic behavior can be
anticipated for all finite systems with an odd number of particles.  In
such cases, the asymptotically dominant contribution to $S_n(t)$ for $n$
even will come from 3-particle configurations where an even particle is
between two odd particles, with all other particles already annihilated.
Conversely, particles with odd labels have a finite probability to
survive indefinitely.  Thus for the finite-particle system $\alpha_n= 0$
for $n$ odd and $\alpha_n=3/2$ for $n$ even.  On this basis, we
anticipate that just two exponents also characterize the individual
particle survival probabilities in the semi-infinite system.

To test these predictions, we performed numerical simulations using two
complementary methods.  The first (naive) approach is to simulate a
suitably-sized system with the initial condition that the right
half-line is completely occupied while the left half-line is empty.
Reflecting boundary conditions are employed at the edges of the system.
The system size is chosen such that the effect of the boundaries is
negligible over the time scale of the simulation.  In the second
approach, particles are created at the right boundary at a rate equal to
the exact time-dependent density of the homogeneous system, $c(t)\simeq
(4\pi t)^{-1/2}$\cite{glauber,spouge}, to mimic the effect of a
semi-infinite system.  This is a more efficient approach, as relatively
long-time simulations can be run on a small systems without being
influenced by boundary effects.  For the survival probability of the
first particle our simulations give $S_1(t)\sim t^{-\alpha_1}$, with
$\alpha_1= 0.225\pm 0.005$.  On the other hand, for the second particle,
$S_2(t)\sim t^{-\alpha_2}$ with $\alpha_2= 0.865\pm 0.015$.  As
anticipated, the survival probability of the first particle ($n=1$)
decays more slowly than $t^{-1/2}$, the particle survival probability in
the bulk.  However, the second particle is much less likely to survive
than a bulk particle.  This arises because the second particle always
has a unique potential left reaction partner, as well as a right
reaction partner.

\begin{figure}
\narrowtext
\epsfxsize=2.5in\epsfysize=2.5in
\hskip 0.3in\epsfbox{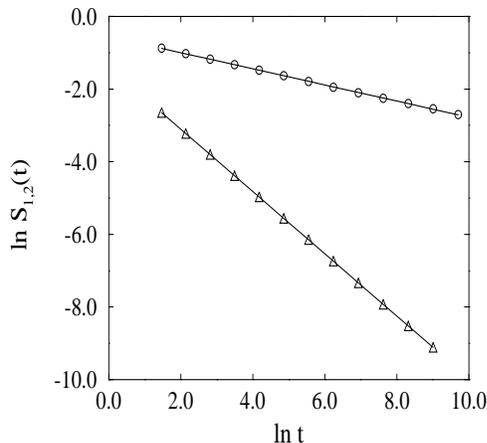}
\vskip 0.15in
\caption{Simulation results for $S_1(t)$ ($\circ$) and $S_2(t)$
($\Delta$).  The quoted exponents are based on the best-fit straight
lines to the data given in the figure.
\label{fig1}}
\end{figure}

To determine whether the survival probabilities $S_n(t)$ are
characterized by only two exponents, we perform a scaling analysis.
Particles far from the interface initially exhibit bulk behavior, where
$S_n(t)\sim t^{-1/2}$ for all $n$.  After a time $t_n\sim n^2$, the
$n^{\rm th}$ particle ``senses'' the interface, and crossover from bulk
to surface kinetics should occur.  Based on the observed asymptotic
behavior of $S_1(t)$ and $S_2(t)$, together with the above crossover
picture, we expect that $S_n(t)$ should exhibit the two distinct scaling
forms for odd and even $n$ respectively,
\begin{equation}
\label{odev}
S_{2n-1}(t)\simeq t^{-1/2}{\cal O}(z),\quad
S_{2n}(t)\simeq t^{-1/2}{\cal E}(z),
\end{equation}
for $t\rightarrow\infty$ and $n\rightarrow\infty$, with $z=nt^{-1/2}$
finite.  Large values of $z$ corresponds to particles sufficiently deep
in the bulk that they are not yet influenced by the heterogeneous
initial condition.  Thus the large argument behavior of the scaling
functions is determined by the survival probability of the homogeneous
system.  Since $S(t)=c(t)\simeq (4\pi t)^{-1/2}$\cite{glauber,spouge}, this
gives ${\cal O}(\infty)={\cal E}(\infty)=1/\sqrt{4\pi}$.  Conversely,
for $z\ll 1$
\begin{equation}
\label{asympt}
{\cal O}(z)\sim z^{\mu_1}, \quad
{\cal E}(z)\sim z^{\mu_2},
\end{equation}
with $\mu_1=2\alpha_1-1$ and $\mu_2=2\alpha_2-1$ to match with the long
time behavior of $S_1(t)$ and $S_2(t)$.  As Figure 2 shows, $S_n(t)$
follows this general description.

\begin{figure}
\narrowtext
\epsfxsize=2.5in\epsfysize=2.5in
\hskip 0.3in\epsfbox{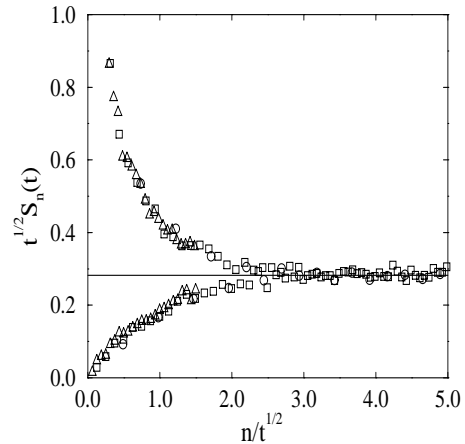}
\vskip 0.15in
\caption{Simulation results for the time dependence of the survival
probability of the $n^{\rm th}$ particle, $S_n(t)$, for $1\leq n\leq
100$.  Shown are the scaling functions ${\cal O}(z)$ (upper curve) and
${\cal E}(z)$ (lower) versus $z=nt^{-1/2}$ at $t=16$ ($\circ$), 256
($\Box$), and 4096 ($\Delta$).  For large $z$, both scaling functions
approach the limiting value $(4\pi)^{-1/2} \approx 0.2821$.
\label{fig2}}
\end{figure}

We can provide a relatively tight rigorous upper bound for the exponent
$\alpha_1$.  We will also present a heuristic argument, based on an
uncontrolled approximation, which turns out to give a relatively
stringent lower bound for $\alpha_1$.  For both situations, our approach
is based on first finding the {\em ultimate} survival probability of the
first particle in a finite-particle system on an infinite lattice and
then using scaling to infer time dependence, from which bounds on
$\alpha_1$ can be inferred.  Let the particles be initially distributed
on $N$ adjacent lattice sites, with $N$ odd.  Ultimately, a unique
particle survives which could be the first, the third, the fifth, {\it
etc.}, in the initial sequence.  Let ${\cal S}_1(N)$ be the probability
that the first particle is this unique survivor.  For $N\to\infty$, we
shall show that this probability scales as
\begin{equation}
\label{s1n}
{\cal S}_1(N)\sim {1\over N^{2\beta_1}}.
\end{equation} 
On the other hand, for $t<N^2$, the finiteness of the system is
immaterial and the survival probability of the first particle should
coincide with $S_1(t)$ in the semi-infinite system.  When $t$ becomes of
the order of $N$, the number of particles remaining will be of order
unity and $S_1(t)$ should ``stick'' at the value ${\cal S}_1(N)$.  Thus
substituting $t^{1/2}$ for $N$ in Eq.~(\ref{s1n}) and equating to
$S_1(t)$ gives $\beta_1=\alpha_1$.

First consider an upper bound for $\alpha_1$.  A naive approximation is
to suppose that every collision between nearest-neighbors has the same
probability to occur.  For the initial $N$ particles there are $N-1$
collision possible and the probability that the first particle survives
after the first collision is $(N-2)/(N-1)$.  This leaves $N-2$ particles
and $N-3$ possible collisions, and the probability that the first
particle survives after this second collision is $(N-4)/(N-3)$.  The
ultimate survival probability of the first particle in this
``democratic'' approximation is
\begin{eqnarray}
\label{lowbound}
{\cal S}_1^D(N)&=&{(N-2)(N-4)\ldots 3\times 1\over (N-1)(N-3)\ldots
4\times 2}\nonumber\\
&\sim& \sqrt{2\over{\pi N}}\quad N\to\infty.
\end{eqnarray}
However ${\cal S}_1(N)\geq {\cal S}_1^D(N)$, because the first particle
has the possibility to ``escape'' on its empty side, and therefore
collisions involving this particle are relatively less likely.  Thus we
conclude\cite{ps}
\begin{equation}
\label{ub}
\alpha_1\leq 1/4.
\end{equation}

While we are unable to obtain a rigorous lower bound for $\alpha_1$, we
have a heuristic approach that gives ${\cal S}_1(N)\sim N^{-\gamma}$,
with $\gamma = 3\sqrt{3}/4\pi$.  This approach is based on first
recasting the annihilation problem into an equivalent aggregation
process\cite{spouge}.  In aggregation, point-like $k$-mers perform
random walks with a mass-independent diffusion coefficient.  When two
polymers of masses $i$ and $j$ happen to occupy the same site, they
irreversibly aggregate into a heavier but still point-like polymer of
mass $i+j$, as represented by the reaction scheme
\begin{equation}
A_i+A_j\rightarrow A_{i+j}.
\end{equation}
To make the connection with annihilation, we categorize polymers
according to whether their mass is odd, $A_o=\{A_1,A_3,\ldots\}$, or
even, $A_e=\{A_2,A_4,\ldots\}$, respectively.  These two classes of
polymers react according to
\begin{eqnarray}
A_e+A_e & \rightarrow & A_e,\nonumber\\
A_o+A_e & \rightarrow & A_o,\\
A_o+A_o & \rightarrow & A_e.\nonumber
\end{eqnarray}
In particular, the parity of odd-mass polymers is not influenced by
even-mass polymers.  Thus by considering only odd-mass polymers,
aggregation is completely equivalent to ARW\cite{spouge}.

\begin{figure}
\narrowtext
\epsfxsize=2.5in\epsfysize=2.5in
\hskip 0.3in\epsfbox{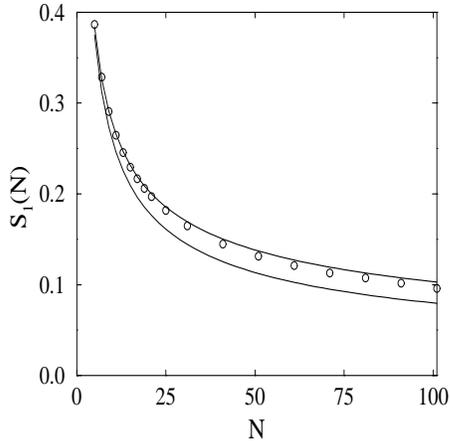}
\vskip 0.15in
\caption{Simulation results for the probability that the
first particle ultimately survives from among an initial group of $N$
adjacent particles (with $N$ odd) on an infinite chain.  The lower curve
represents the rigorous bound given in Eq.~(4), and the upper curve
represents the heuristic bound from Eqs.~(8) and (11).
\label{fig3}}
\end{figure}

If one associates the initial particles in ARW with monomers in
aggregation, then the first particle survives in the ARW process if the
mass of the leftmost polymer in the corresponding aggregation process
remains odd throughout the evolution.  With this equivalence, we now
postulate, in the spirit of a Kirkwood approximation, that ${\cal
S}_1(N)$ obeys the recursion relation
\begin{equation}
\label{recura}
{\cal S}_1(N+2)\approx{\cal S}_1(N)F(N+2)
\approx\prod_{\stackrel{k=1}{k~ {\rm odd}}}^{N+2} F(k).
\end{equation}
Here $F(k)$ is the probability that the first collision between three
random walks, which are initially at $x_0=1$, $y_0=k-1$, and $z_0=k$,
occurs between the particles at $k-1$ and $k$.  While this approximation
is uncontrolled, it turns out to give a relatively stringent upper bound
for the true behavior of ${\cal S}_1(N)$ (see Fig.~(\ref{fig3})).

To compute $F(k)$, we map the problem of three random walks, initially
at $x_0$, $y_0$, and $z_0$ in one dimension, onto a single random walk
in three dimensions\cite{fg,dba}.  A collision between particles 1 and
2, and between 2 and 3 imposes the boundary conditions that the
probability distribution vanishes when $x=y$ and $y=z$,
respectively.  This implies that the three-dimensional walk is confined
to the wedge-shaped region defined by $x\leq y$ and $y\leq z$.
Using image techniques, the probability distribution in the continuum
limit can be written down, from which the desired eventual collision
probability follows after some tedious calculation.

For simplicity, we give an alternative derivation which exploits the
isomorphism between the eventual collision probability and
electrostatics\cite{harmonic}.  The three-dimensional region $x\leq
y$ and $y\leq z$ can be projected into two dimensions, with the
allowed region now a wedge of opening angle $\Omega=\pi/3$\cite{fg,dba}.
Any initial state of three random walks maps to a point in this
two-dimensional domain.  To determine the co-ordinates of this initial
point, we need to define a two-dimensional co-ordinate system which is
perpendicular to the axis $\hat e_1= (1,1,1)/\sqrt{3}$ generated by the
intersection of the planes $x=y$ and $y=z$.  A convenient basis
is $\hat e_2=(0,-1,1)/\sqrt{2}$ and $\hat e_3=(-2,1,1)/\sqrt{6}$.  The
initial condition $\vec r_0=(x_0,y_0,z_0)$ has components $d_2=\vec
r_0\cdot\hat e_2=(z_0-y_0)/\sqrt{2}$ and $d_3=\vec r_0\cdot\hat
e_3=(-2x_0+y_0+z_0)/\sqrt{6}$ in the $\hat e_2$ and $e_3$ basis.  Within
the two-dimensional wedge, with the horizontal axis defined as the locus
where $y=z$, an arbitrary initial condition corresponds to a
horizontal displacement of $(z_0+y_0-2x_0)/\sqrt{6}$ and a vertical
displacement of $(z_0-y_0)/\sqrt{2}$.  Thus the initial point is
inclined at an angle
\begin{equation}
\label{angle}
\theta=\tan^{-1}{d_2\over d_3}= 
\tan^{-1}\left[\sqrt{3}\left({{z_0-y_0}\over{z_0+y_0-2x_0}}\right)\right]
\end{equation}
with respect to the horizontal.

We want to compute the probability that the random walk eventually hits
the horizontal axis, corresponding to particles 2 and 3 colliding.  From
the isomorphism with electrostatics, this probability equals the
integral of the electric field over the horizontal axis, which is
generated by the unit charge at the initial position of the random
walk\cite{harmonic}.  To simplify computation of this integral, perform
the conformal transformation $w=z^3$ to open the wedge onto the upper
half plane, so that the initial point is now inclined at an angle
$3\theta$ with respect to the positive real axis.  In real co-ordinates,
for an initial point at $(x_0,y_0)$ with a grounded plane at $y=0$, the
electric field at $(x,0)$ equals
\begin{equation}
\label{field}
{y_0\over\pi}{1\over{(x-x_0)^2+y_0^2}}.
\end{equation}
The integral of this field over any interval on the $x$-axis gives the
probability that a random walk which starts at ($x_0,y_0$) eventually
hits this interval.  This gives ${\phi\over \pi}$, where $\phi$ is the
angular size of the interval as seen from the location of the charge.
In our case the appropriate interval $x=(0,\infty)$ has angular size
$\phi=\pi-3\theta$, so that
\begin{eqnarray}
\label{theta}
F(k)&=&1-{3\theta\over\pi}\nonumber\\
&=&1-{3\over \pi}\tan^{-1}{\sqrt{3}\over 2k-3}\nonumber\\
&\to&1-{3\sqrt{3}\over 2\pi k}, \qquad k\to\infty,
\end{eqnarray}
where the second line is obtained by the substitution of the initial
condition $x_0=1$, $y_0=k-1$, and $z_0=k$, as specified by the
definition of $F(k)$. Note that this result is easily generalizable to
the case where the three particles have distinct diffusivities.

Using Eq.~(\ref{recura}), we obtain
\begin{equation}
\label{prod}
{\cal S}_1(N)\approx \prod_{\stackrel{k=1}{k~ {\rm odd}}}^{N} 
\left(1-{3\sqrt{3}\over{2\pi k}}\right)\sim N^{-3\sqrt{3}/4\pi}.
\end{equation}
Thus we arrive at the following approximate expression for $\alpha_1$
\begin{equation}
\label{lb}
\alpha_1= {3\sqrt{3}\over 8\pi}\simeq 0.2067\ldots
\end{equation}

Let us now consider the exponent $\alpha_2$.  We write the survival
probability of the second particle as $S_2(t)=P(t)S_R(t)$, where
$P(t)\sim t^{-1/2}$ is the probability that the particle has not been
annihilated by its single left neighbor and $S_R(t)$ is the probability
that the particle has not been annihilated by any particle to its right.
This latter probability decays as $t^{-1/4}$, since in the bulk the
survival probability $S(t)=S_L(t)S_R(t)=S_R(t)^2$ varies as $t^{-1/2}$.
This approximation immediately leads to $\alpha_2=3/4$ which can be
expected to be the lower bound.  Unfortunately, we have been unable to
construct a non-trivial upper bound for $\alpha_2$.  A trivial upper
bound, however, is provided by the survival probability of the central
particle in a 3-particle system.  Consequently, we have the bounds
\begin{equation}
\label{lower}
{3\over 4}<\alpha_2<{3\over 2}.
\end{equation}

We now turn to a related and useful microscopic characterization of the
reaction process, namely, the probability that a particle is eventually
annihilated by its n$^{\rm th}$ nearest-neighbor, $P(n)$\cite{serial}.
For homogeneous reaction processes, $P(n)$ typically decays as a power
law in $n$, $P(n)\sim n^{-\psi}$, and the exponent $\psi$ is related to
the time dependence of the survival probability.  For homogeneous ARWs,
for example, the particle survival probability $S(t)$ decaying as
$t^{-\alpha}$ (with $\alpha=1/2$) implies that the probability that a
particle is annihilated at time $t$ is $-{dS(t)\over{dt}}\sim
t^{-\alpha-1}$.  The annihilation probabilities for given $t$
and $n$ can now be related by
\begin{equation}
\label{relate}
-{dS(t)\over{dt}}\,dt = P(n)\,dn.
\end{equation}
For diffusive transport, $n$ scales as $t^{1/2}$ and Eq.~(\ref{relate}),
together with the defining relations for $\alpha$ and $\psi$, then gives
$\psi=1+2\alpha$.

\begin{figure}
\narrowtext
\epsfxsize=2.5in\epsfysize=2.5in
\hskip 0.3in\epsfbox{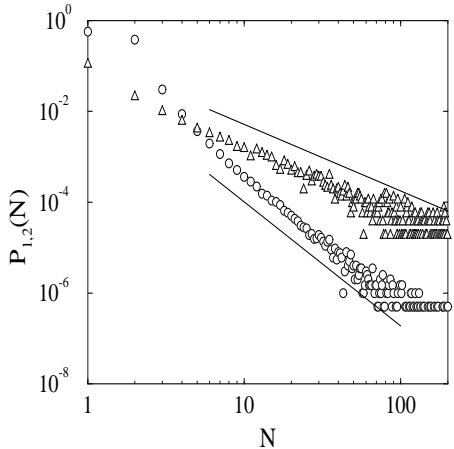}
\vskip 0.15in
\caption{Simulation results for the annihilation probabilities 
$P_1(n)$ ($\circ$) and $P_2(n)$ ($\Delta$).  For reference, the straight
lines have slopes $-1.45$ and $-2.73$.
\label{fig4}}
\end{figure}

We now apply the same line of reasoning for the semi-infinite system.
Let us define the probability that the first particle is annihilated by
the $n^{\rm th}$ particle as $P_1(n)\equiv n^{-\psi_1}$; our
previous estimate for $\alpha_1$ gives $\psi_1=1+2\alpha_1\approx 1.45$.
Similarly for the second particle, $P_2(n)\equiv n^{-\psi_2}$, with
$\psi_2=1+2\alpha_2\approx 2.73$.  As shown in Fig.~\ref{fig4}, these
expectations are consistent with our data.  One additional interesting
feature is that the second particle annihilates with the first particle
with probability $\approx 0.5704$, while it annihilates with any other
particle with probability $\approx 0.4296$.

Finally, we study the spatial distribution of the leftmost particle.
This refers to the extreme particle that currently exists and is not
necessarily the initial particle that was leftmost.  To provide some
perspective on the behavior one might anticipate, first consider this
question for two simpler systems with the same semi-infinite
concentration profile, $c(x,t=0)=\Theta(x)$.  For freely diffusing
particles, the long time spatial distribution in the continuum limit is
the error function, $c(x,t)={1\over 2}[1+{\rm erf}(x/\sqrt{4t})]$.  From
this, the concentration at $x=0$ remains fixed at the value 1/2, while
the typical position of the leftmost particle is $x_-\approx -\sqrt{t\ln
t}$.  Thus freely diffusing particles substantially penetrate the
negative half-line.  For coalescing random walks, where particles react
by $A+A\to A$, the leftmost particle simply undergoes free diffusion.
Thus the concentration at $x=0$ vanishes as $t^{-1/2}$, while the
typical position of the leftmost particle remains at the origin.
 
\begin{figure}
\narrowtext
\epsfxsize=2.5in\epsfysize=2.5in
\hskip 0.3in\epsfbox{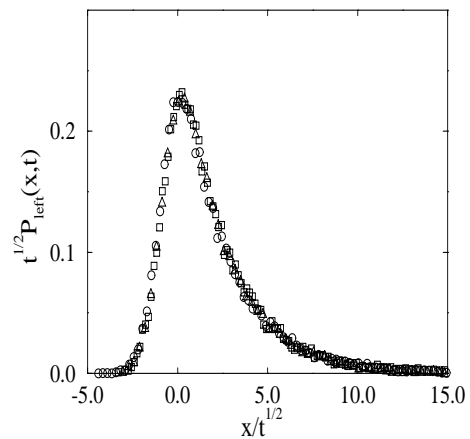}
\vskip 0.15in
\caption{Simulation data for the probability distribution of the
location of the leftmost particle.  Shown is the scaled distribution
$t^{1/2}P_{\rm left}(x,t)$ versus scaled co-ordinate $x/t^{1/2}$ for
$t=16$ ($\circ$), 256 ($\Box$), and 4096 ($\Delta$).
\label{fig5}}
\end{figure}

For ARWs, the concentration profile can be readily computed for
arbitrary initial conditions from the direct correspondence to the
Glauber solution of the kinetic Ising model\cite{glauber}.  This
calculation gives
\begin{eqnarray}
\label{exact}
c(n,t)&=&{\rm e}^{-2t}\left[\sum_{m=0}^\infty (-1)^n I_{n+m}(t)
\sum_{m=1}^\infty I_{n-m}(t)\right.\nonumber\\
\medskip
& &\qquad\qquad \left.+\sum_{m=0}^\infty I_{n+m}^2(t)\right].
\end{eqnarray}
This profile exhibits an overall $t^{-1/2}$ decay of the density and a
non-trivial spatial dependence.  In the long-time limit, the above
expression can be reduced to the scaling form
\begin{equation}
\label{cnt}
c(n,t)=t^{-1/2}{\cal C}(z),
\end{equation}
with $z=nt^{-1/2}$ and where the scaling function ${\cal C}(z)$ can be
expressed in terms of single and double integrals of the error
functions.  While the computation of the density profile requires the
knowledge of the two-point correlation function of the equivalent
kinetic Ising model, the spatial distribution of the leftmost particle
would require the knowledge of all the $n$-point correlation
functions. Although it is in principle possible to obtain such functions
\cite{bed}, this is a considerable analytical task and we merely use
simulations to provide numerical data for the spatial distribution of
the leftmost particle, $P_{\rm left}(x,t)$ (Fig.~\ref{fig5}).  This
distribution obeys the expected scaling behavior and is asymmetric in
character, with the negative-$z$ tail decaying as $e^{-z^2}$ while the
positive-$z$ tail decays as $e^{-z}$ with $z=xt^{-1/2}$.  From this
data, we find, for example, that the average position of the leftmost
particle varies as $x_{\rm left}(t)\approx 1.7\, t^{1/2}$.

In summary, we studied basic properties of a semi-infinite population of
annihilating random walks near a free interface.  Since particles near
the interface have fewer potential reaction partners than bulk
particles, these interface particles should be more long-lived than
those in the bulk.  This naive expectation turns out to be only
partially correct.  For the $n^{\rm th}$ particle from the interface
(with $n=1$ corresponding to the particle at the interface), the
survival probability $S_n(t)$ decays as $t^{-\alpha_n}$, with
$\alpha_n\approx 0.225$ for all odd values of $n$, but $\alpha_n\approx
0.865$ for all even values of $n$.  These exponents can be viewed as
characterizing the surface critical behavior of ARWs in one dimension.

This alternating behavior stems from the fact that an odd particle can
eventually become the leftmost particle in the system and hence be
long-lived.  Conversely, an even particle will always have potential
reaction partners on both sides and therefore has a relatively shorter
lifetime.  For the odd particles, the bounds $3\sqrt{3}/8\pi\leq
\alpha_1\leq 1/4$ were derived, with the  lower bound non-rigorous but
numerically accurate, by considering the eventual survival probability
in a finite group of $N$ particles, with $N$ odd.

The relative longevity of the interface particle is also reflected in
the fact that the mean position of its reaction partner drifts slowly to
the right as $1.7\,t^{1/2}$.  The functional form of the probability
distribution of $x_{\rm left}$ could be obtained, in principle, from the
$n$-point correlation functions of the equivalent kinetic Ising model;
this appears to be a formidable and unenlightening task.  The numerical
data for $P_{\rm left}(x,t)$ clearly exhibits scaling and shows that the
position of the leftmost particle is described by a single length scale
which varies as $\sqrt{t}$.

We thank E.~Ben-Naim, B.~Derrida, and F.~Leyvraz for helpful
discussions.  This research was supported in part by the Swiss National
Foundation, the ARO (grant DAAH04-96-1-0114), and the NSF (grant
DMR-9632059).

\end{multicols} 
\end{document}